\documentstyle[pra,amssymb,aps,multicol,epsfig]{revtex}%multicol,ifthen,epsfig]{revtex}
\newcommand{\ket}[1]{\left| {#1} \right\rangle}
\newcommand{\bra}[1]{\left\langle {#1} \right|}
\newcommand{\op}[1]{\hat{#1}^{}{}}
\newcommand{\tr}[1]{\mbox{Tr}\left[{#1}\right]}
\newcommand{\half}{\frac{1}{2}}
\newcommand{\Half}{{1/2}}
\newcommand{\ds}{\displaystyle}
\begin{document}
%======= TITLE =========================================
\title{Information transfer and fidelity in quantum copiers}
\author{P. Deuar\cite{PDemail} and W. J. Munro}
\address{Centre for Laser Science,Department of Physics, University of
Queensland, QLD 4072, Brisbane, Australia}
\date{\today}
\maketitle
%=======ABSTRACT=========================================
\begin{abstract}
  We find that very different quantum copying machines are optimal depending
on the indicator used to assess their performance. Several quantum copying
machine models acting on nonorthogonal input states are investigated, and
assessed according to two types of criteria: transfer of (Shannon) information
encoded in the initial states to the copies, and fidelity between the copies and
the initial states. Transformations that optimize information
transfer for messages encoded in qubits are found for three
situations: (1) when the message is decoded one state at a time; (2)
with simple schemes that allow  the message to be encoded using
block-coding schemes; and (3) when the
copier produces independent copies. If the message is decoded one symbol
at a time, information is best copied by a Wootters-Zurek copier.
\end{abstract}
\pacs{03.67.Hk, 03.65.Bz}

\begin{multicols}{2}
\narrowtext
%======INTRODUCTION======================================
\section{Introduction}
\label{INTRO}
  Quantum copying has attracted considerable interest in recent
years, ever since the discovery of the no-cloning theorem\cite{WZ:82,Dieks:82},
and the universal quantum copying machine\cite{BuzekH:96} which copies arbitrary unknown
qubits with the best fidelity.
To date, most treatments have used fidelity to characterize the quality of the
copies produced. The fidelity between two quantum states characterized by density
operators $\op{\rho}_1$ and $\op{\rho}_2$ is 
\begin{equation}
  F(\op{\rho}_1, \op{\rho}_2) = \left\{\tr{\sqrt{(\op{\rho}_1)^{\Half}\op{\rho}_2(\op{\rho}_1)^{\Half}}\,\,}\right\}^2.
\end{equation}
  A good summary of its properties is given in Ref.\cite{Jozsa:94}.
  In the case where one of the states is pure, the fidelity is simply the
square of the overlap between the two states.

Many authors\cite{Barnumetal:96,GisinH:97,GisinM:97,Brussetal:98,Gisin:98,DuanG:98b,Cerf:98,Buzeketal:98a,Werner:98,Buzeketal:98b,Muraoetal:99} have made use of two fidelity measures for quantum copiers:
the {\it global fidelity} of the combined output (both copies) of the copier,
with respect to a product state of (unentangled) perfect copies,
 and the {\it local fidelity} of one
copy with respect to the original input state.
Here, we will concentrate on a different indicator of copying success:
mutual information content between the copies and the originals. One
finds that which copier is optimal depends greatly on which indicator
is used. In practice, this will mean that what sort of quantum copier
is best depends on what one wants to do with the copies afterward.

  This article proceeds in the following fashion:
After commenting on some drawbacks of fidelity, and why one might want
to use different indicators, we outline exactly what we mean by
information content between copies and originals in Sec.~\ref{F&I}.
General features of the copiers that will be considered are mentioned in
Sec.~\ref{GPROP}. Copiers optimized for maximum copied information
are given in Sec.~\ref{3COP} (and derivations are given in
Appendixes~\ref{APP} and~\ref{NOEAPP}) for three cases: (1) when the information is
decoded from the copies one state at a time; (2) when efficient block-coding
schemes are used to transmit as much information as is allowed by the
Holevo bound; and (3) when the copies are an unentangled product state.
In Sec.~\ref{COMP} the performance of these copiers is assessed according to information transfer
and fidelity criteria, and compared to the performance of fidelity-optimized
copiers known previously.

%========== FIDELITY AND MUTUAL INFO ==================================
\section{Mutual Information and Fidelity Measures}
\label{F&I}
%========== FIDELITY ==================================================
\subsection{Fidelity, and some of its drawbacks}
\label{F}
  Fidelity is used in many fields as an indicator of closeness between
two states, and is often quite useful.  It is probably also one of the easiest such indicators to
calculate. However, it sometimes suffers from a number of drawbacks
(examples of which are given below) when
used as a measure of closeness over broad classes of
systems, so there will be times when one wants to use a different
indicator. 

   While a fidelity of $1$ obviously implies identical states,
and $0$ implies orthogonal states, what intermediate values mean is
highly dependent on the particular states that are being
compared, particularly if both states are impure. Thus a
statement such as ``The fidelity between the two states was $x$,'' to be unambiguous, often
needs considerable additional information on the states that were
compared. To give an example:
For standard optical coherent states
of complex amplitude $\alpha$, given by 
\begin{equation}
\ket{\alpha} = e^{-(\Half)|\alpha|^2}\sum_{n=0}^{\infty}
\frac{\alpha^{n}}{n!} \ket{n},
\end{equation}
the fidelity between two pure coherent states $\ket{\alpha}$ and
$\ket{\alpha+1}$ is always constant:
\begin{equation}
F(\ket{\alpha}\bra{\alpha},\ket{\alpha+1}\bra{\alpha+1}) = \frac{1}{e}.
\end{equation}
Now if \mbox{$\alpha = 0$}, the two states are the vacuum and a
low-photon-number coherent state --- states with qualitatively
different properties. However,  if $\alpha$ is large, then
$\ket{\alpha}$ and $\ket{\alpha+1}$ are macroscopic, and
experimentally indistinguishable, but the fidelity between them is
still $1/e$.

  Another drawback of fidelity is that it is not directly related to
other quantities commonly measured in experiments. While the fidelity
is an expectation value of an observable (the observable being
either one
of the two states), it cannot usually be calculated from the results of
experiments whose aim is to do something other than measure
fidelity. It is not in general directly related to expectation values
or measurement probabilities of other quantities, so it does not say
much about the usefulness of a copy. In this sense, fidelity 
characterizes the closeness of the mathematical representation of 
physical states more than the closeness of the physical properties of
those states.
Of course, in many
situations, these
two types of closeness are equivalent, but not always.  

 For the specific case of quantum copiers, 
global or local fidelities are not robust to unitary transformations
made on the copies individually after all copying has been completed,
and also can be very high even though the copies are uncorrelated with
the originals. 
For example, suppose a message is encoded in a binary alphabet of
orthogonal states \mbox{$\ket{0},\ket{1}$}, and sent through a lossless
communication channel that interchanges the states, i.e., they undergo the
transformation
\begin{eqnarray}
\ket{0} & \to \ket{1},\\
\ket{1} & \to \ket{0}, 
\end{eqnarray}
then the fidelity of the transmitted with respect to the initial state
is {\it zero}, but nothing of interest has been lost. It is
sufficient for an observer receiving the message to relabel the states
which they receive to recover the original message.

Conversely, consider the situation where very nonorthogonal states $\ket{a}$
and $\ket{b}$ are used to encode a message. Using appropriate
error-correction schemes, some information can be reliably transmitted
with this encoding. However, now suppose that the message is intercepted by an
eavesdropper, who simply sends the same state
\mbox{$\sqrt{1/2}\,(\ket{a}+\ket{b})$} on to the intended receiver
every time. The fidelity between sent and received states is still
very high, but the received message carries no information from the sender.

Global fidelity measures are often particularly removed from experimental
results, since they compare the combined state of both copies with a
perfect copy state that is generally unattainable due to the no-cloning
theorem. However, in practice, one usually makes copies so they can subsequently
be considered only individually. 

%========= MUTUAL INFO MEASURES ==========================================
\subsection{Mutual information measures}
\label{IM&IH}
   A different, natural,  measure of copying efficiency that can be used is the
amount of mutual (Shannon) information\cite{Shannon:48} shared between the original
states, and the copies. 
This mutual information does away with some
of the drawbacks of fidelity, as discussed below.
  
Consider two observers: one of them, the sender (labeled $A$),  is sending states chosen from some
ensemble, where the {\it a priori} probability of sending the $i$th variety of state
is $P^A_i$. 
The other observer, the receiver (labeled $B$), makes
measurements on one of the copies, obtaining the $j$th measurement
result with probability $P^B_{j|i}$, given that the $i$th
state was sent into the copier. The amount of information (in bits per
sent state) that the receiver has obtained from the sender is the
Shannon mutual information, given by 
\begin{equation}\label{IAB}
  I(A:B) = \sum_{i,j}  P^A_i P^B_{j|i} \log_2 \frac{P^B_{j|i}}{P^B_j},
\end{equation}
where $P^B_j$ is the overall probability of the receiver
obtaining the $j$th measurement result, averaged over the input states.

To use this measure to characterize a copying machine,
rather than the specific message encoding or the ingenuity of the
receiver in constructing a measuring apparatus, three points should be
noted.
First, even if a perfect copier is used, the amount of information
that can be transmitted from originals to copy depends on the ensemble of
states that is used to encode the message. Thus, the information about
the original extractable from the copy $I(A:B)$ must be compared
to the amount of information extractable from the original $I(A:A)$. 

Secondly, if observer $B$ makes a suboptimal (in terms of recovering
the original message) set of measurements, then $B$'s stupidity will
affect the mutual information. 
To eliminate the effect of $B$'s ingenuity (or lack of it), it has to be
assumed that optimal measurements are made to recover the encoded
message.  

Thirdly, 
a characterisation of the copier would usually involve examining its
information-copying performance for a given set of input
states. However, these may occur with various {\it a priori}
probabilities $P_i^A$. We will take the case where these probabilities are
chosen to encode the maximum amount of information in the signal
states to be most representative of the behavior of information in
the copier. 
Thus the mutual information quantities that will be used in later
sections of this article are $I_m(A:B)$ and $I_m(A:A)$, given by
\begin{equation}\label{Im=}
I_m(A:B) = \max_{\left\{P^A_i\right\}} \left[ \max_{\left\{\cal{E}_B\right\}}{I}(A:B)\right],
\end{equation}
where $\left\{P^A_i\right\}$ denotes the set of {\it a priori}
probabilities of $A$ using the $i$th state in the encoding of the
message, and  $\left\{\cal{E}_B\right\}$ is the set of all positive valued
operator measures\cite{POVMexplanation}.
  We will call $I_m$ the {\it copied information}.

While this quantity can be more laborious to calculate, it has some
advantages over fidelity. It is unchanged by relabeling or by
local unitary
transformations on the copies after they have left the copier, as well
as always being zero if the copies are independent of the originals.

Also,
 such mutual information is a physical quantity of
interest in its own right, and is in fact what one is interested in in
many fields (such as cryptography, for example). Even where
this is not the case, mutual information between originals and copies
can often be calculated from probability distributions of experimental
measurements. Furthermore, it is clear what the statement  ``the
mutual information transfer from $A$ to $B$ is $x$\,'' means physically,
with no further knowledge of the actual quantum states that were
sent.
  It could be said that the information-copying capacity of a quantum cloner quantifies the
{\it practical usefulness}, in many situations, of the copies
 produced by it.

There is a qualitative difference between information-theoretic
quantities such as copied information, and quantities such as fidelity.
 Fidelity, and similar quantities such as the Hilbert-Schmidt norm or
the Bures distance, 
are quantifications of relations between two quantum states (or, more
precisely, between their mathematical representations), while
information-theoretic quantities deal with the relations between
{\it ensembles} of states. This is the reason that they are robust to
such postcopying effects as relabeling of the copy states.

%============ IH & IM ============================================
\subsection{Ultimate and one-state copied information}
\label{IH&I1}
    Consider the situation discussed in the previous
subsection. Observer $A$
encodes a message into a sequence of quantum states, chosen from a set
of states \mbox{$\left\{\op{\rho}^A_i \right\}$} labeled by the index
$i$. 
Each of the sent 
states has an {\it a priori} probability
$P^A_i$ of being the $i$th one in the set. When the copying machine
acts on the signal state 
$\op{\rho}^A_i$ , it produces a copy state
 $\op{\rho}^B_i$, which is usually
different from the original. 
It has been shown\cite{Holevo:73,CavesD:94} that the mutual
information between $A$ and $B$ can be no more than $I_H(A:B)$, given by
\begin{mathletters}\label{Holevob}\begin{eqnarray}
I_m(A:B) &\le& I_H(A:B)\nonumber\\ &=& S\left(\sum_i P^A_i \op{\rho}^A_i \right) -
\sum_i P^A_i S\left(\op{\rho}^A_i \right), 
\end{eqnarray}
where $S(\op{\rho})$ is the von Neumann quantum entropy of state
$\op{\rho}$:
\begin{equation}
S(\op{\rho}) = -\tr{\ \op{\rho} \log_2 \op{\rho}\ },
\end{equation}\end{mathletters}
a result known as the Holevo theorem.
  
In practice, the transmitted information will usually be significantly less
than $I_H(A:B)$. However, it has been
shown\cite{Hausladenetal:96,Holevo:98} that if $A$ encodes the
message using only certain sequences of states out of all the possible
ones (although still respecting the {\it a priori} probabilities of
individual states), and $B$ makes measurements on whole such sequences
rather than on individual states, then as the length of these sequences
increases, the information capacity per state can approach arbitrarily
close to the Holevo bound $I_H(A:B)$.  This is called a block-coding
scheme, and such a  communication setup is analogous to sending
and distinguishing
only whole ``words'' at a time in the message, rather than individual
``letters.''  In this analogy, letters correspond to individual quantum
states, and words to sequences of them. Naturally, only special
choices of the ``words'' to be used will
approach the Holevo bound, Eq.\ (\ref{Holevob}).  

With this in mind, there are two obvious candidates for a mutual
information quantity with which to
characterize copiers: the {\it ultimate copied information} given
by $I_H$, and the {\it one-state copied information} $I_1$,
which is the maximum information obtainable if measurements are made on only
one state at a time. Both will be considered in what follows.

%====== GENERAL PROPERTIES =========================================
\section{General Properties of the Copying Setups Considered}
\label{GPROP}
  In the interest of clarity and simplicity (and, one must admit, ease  of
analysis), only the most basic relevant copying setups have been
investigated. This should make the principles involved easier to see,
without introducing too much complexity.

  Thus, we will consider the case where observer $A$ encodes a message
into a binary sequence of pure quantum states
\mbox{$\op{\rho}^A_i = \ket{\psi^A_i}\bra{\psi^A_i}\,\,\, (i=1,2)$} 
with equal {\it a priori}
probabilities of being sent \mbox{($P_i^A = \half$)}.  The $P^A_i$ are
chosen to be one-half for two reasons: (1) this is the simplest case; (2)
this is the situation where the maximum amount of information is
encoded in the input states.  

Since there are only two input states, the dimension of the relevant
Hilbert space can be reduced to 2 by appropriate unitary
transformations, because the states span at most a two-dimensional
manifold in Hilbert space. Any such can be written (discarding an
irrelevant phase factor) in an orthogonal basis \mbox{$\left\{
\ket{+}, \ket{-}\right\}$} as 
\begin{mathletters}\begin{eqnarray}
\ket{\psi^A_1} &=& \cos\theta\ket{+} + e^{i\mu}\sin\theta\ket{-}, \\
\ket{\psi^A_2} &=& \sin\theta\ket{+} + e^{-i\mu}\cos\theta\ket{-},
\end{eqnarray}\end{mathletters}
where the parameter $\theta$ ranges from $0$ to $\pi/4$ 
(other values of $\theta$ are equivalent to a
relabeling of the two states). In the rest of the article, $\mu$
will be taken to be zero for simplicity, although all results can
easily be extended to the nonzero case. 
This, then, gives a one-parameter family of input states:
\begin{mathletters}\label{instates}\begin{eqnarray}
\ket{\psi^A_1} &=& \cos\theta\ket{+} + \sin\theta\ket{-}, \\
\ket{\psi^A_2} &=& \sin\theta\ket{+} + \cos\theta\ket{-}.
\end{eqnarray}\end{mathletters}
These can be fully labeled by the fidelity between them,
\begin{equation}
f = F(\op{\rho}^A_1, \op{\rho}^A_2) = \sin^2(2\theta).
\end{equation}

In similar fashion, by taking the least complex case, the copiers considered will be unitary, create
only two copies, and be symmetric. By symmetric we mean that the reduced
quantum states of both copies by themselves are equal.

The unitarity of the copying process implies a ``black box'' process: no external
disturbance is required during the copying. Probabilistic
copiers\cite{DuanG:98b,DuanG:98a} are not considered here. 

Physically, there are two subsystems $o$ and $c$ (which can be
considered two dimensional
for reasons outlined above) put into the
unitary copying machine, and two come out. At the input, the subsystem
$o$ contains the original state to be copied, while $c$
contains a ``blank'' state that is always the same, irrespective of
what enters at $o$. Both subsystems contain the (usually
imperfect) copies when they exit the copier, while an ancillary machine state
subsystem ($x$) is also used in some of the copiers. At the input, all
three subsystems are unentangled, while at the output, entanglement is
usually present.
Due to unitarity, the full entangled output states
consisting of all three subsystems \mbox{$o$, $c$, and $x$} are pure, but the
states of individual subsystems are in general mixed.

%================== COPIERS ===============================================
\section{Three Information-Optimized Quantum Copiers}
\label{3COP}
    In this section, we present transformations for several copiers optimized
for information transfer to the copies, given a binary sequence of
equiprobable input states. All these copiers are symmetric. The input
states are in general nonorthogonal, and the degree of orthogonality
is characterized by $f$, the square of the overlap between the two
input states $\op{\rho}^A_1$ and $\op{\rho}^A_2$.
 These will be compared to known fidelity-optimized
copiers in the next section.

%================== WZ =====================================================
\subsection{Copiers that optimize the one-state copied information}
\label{WZ}
Rather than carry out a tedious optimisation, it stands to reason that
if any unitary copier allows one to extract as much information about the originals
from the copies as from the originals themselves, then it achieves the optimum.
 Is there such a copier?

Perhaps surprisingly, one finds that the Wootters-Zurek (WZ) quantum copying
machine\cite{WZ:82,BuzekH:96} (used in the original proof of the 
no-cloning theorem) allows one to extract as much information (using a
one state at a time extraction) from
either of the copies as from the original.  One can imagine that the
same information transfer could be achieved by making measurements on
the originals, and sending the results classically, but that a simple unitary
transformation with no coupling to the external environment can achieve the same is perhaps less obvious.
What is
more, the WZ copier does much  better than any
fidelity-optimized copiers, as will be seen later.

Explicitly, the transformation of the input states\ (\ref{instates}) is given
by
\begin{mathletters}\label{WZtransf}\begin{eqnarray}
\ket{\psi^A_1} &\to& \sin\theta\ket{++} + \cos\theta\ket{--}, \\
\ket{\psi^A_2} &\to& \cos\theta\ket{++} + \sin\theta\ket{--}, 
\end{eqnarray}\end{mathletters}
where the basis vectors $\ket{+-}$, etc., indicate tensor products \mbox{$\ket{+}_o\ket{-}_c$} of
the basis vectors for the $o$ and $c$ copy subsystems, respectively.
The combined state of the copies is highly entangled, but
the reduced density matrices of the copies (the full output density matrices traced over all subsystems except one copy)  are
in the classically mixed states
\begin{mathletters}\begin{eqnarray}
\op{\rho}^B_1 &=& \left(\begin{array}{cc}\cos^2\theta&0\\0&\sin^2\theta\end{array}\right),\\
\op{\rho}^B_2 &=&  \left(\begin{array}{cc}\sin^2\theta&0\\0&\cos^2\theta\end{array}\right).
\end{eqnarray}\end{mathletters}

The one-state copied information, which is the same as can be
extracted from the originals,  is
\begin{mathletters}
\begin{equation}\label{WZIm}
I_1^{\text{WZ}} = \half\left[(1+q)\log_2(1+q) + (1-q)\log_2(1-q)\right],
\end{equation}
where $q$, which we will call  the distinguishability parameter, is
\begin{equation}
  q = \sqrt{1-f}.
\end{equation}\end{mathletters}
From the purely classical nature of $\op{\rho}^B_i$, it follows that the
ultimate copied information $I_H^{\text{WZ}}$ is no bigger than $I_1^{\text{WZ}}$.
In fact, applying more WZ copying machines to the copies made by the first one,
in a cascade effect, creates larger
numbers of copies, each of which still carries the same amount of (one-state)
information as the original message.
 In this way, arbitrary
numbers of optimal copies can be made --- similarly to how one can make arbitrary
numbers of copies of classical information.

The local fidelity between a copy and the originals is
\begin{equation}
F(\op{\rho}^A_i, \op{\rho}^B_i) = 1-\frac{f}{2}.
\end{equation}
   
There are other copiers related to the WZ copier which allow the same
optimal one-state information transfer. One example is the family of
copying transformations created by applying identical local unitary
transformations on both copies after they come out of the
WZ copier. The particular transformation presented above in Eq.\ (\ref{WZtransf})
is the one that gives the best local fidelity out of this family
of transformations.

%===============  HOLEVO ================================================
\subsection{Copiers without ancilla that optimize the ultimate copied information}
\label{HOLEVO}

   It is also of interest how well information can be transmitted
when the possibility of complicated block-coding schemes is allowed,
   as discussed in Sec.~\ref{IH&I1}.
To make the calculations relatively tractable analytically, we have
made two restrictions on the copiers that we considered for this task.

\begin{figure}
\center{\epsfig{figure=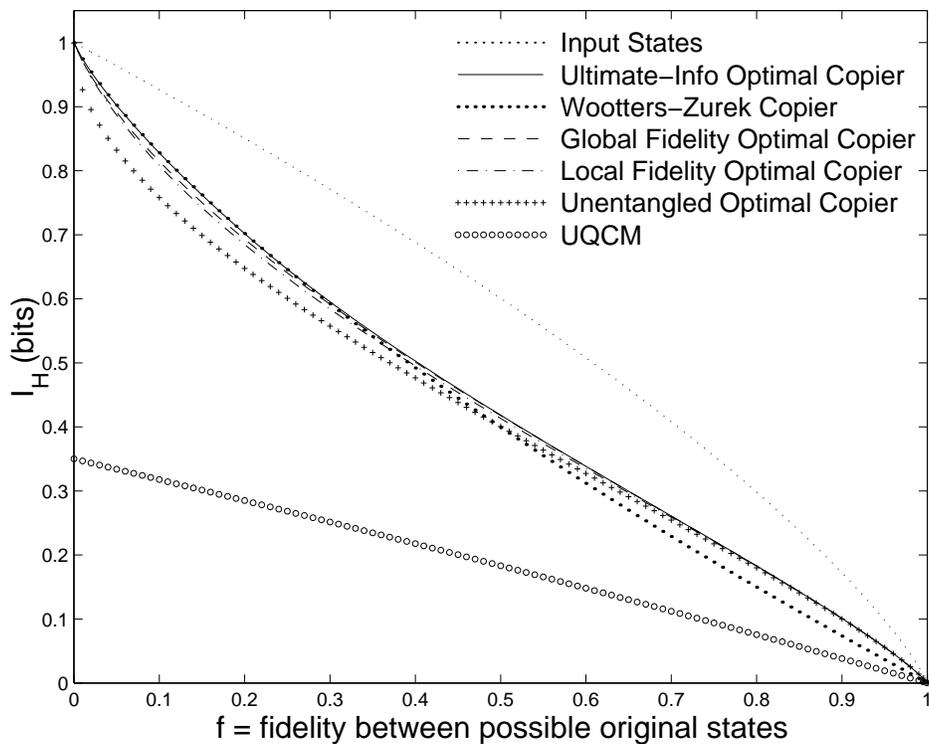,width=125mm}}
\caption{
   Ultimate (Holevo bound) copied information (in bits per signal state) $I_H$ for the copying
   machines discussed in Secs.~\ref{3COP} and \ref{COMP} and Appendix~\ref{FAPP}, depending on the
   fidelity $f$ between the two pure input signal states. The Holevo
   bound on information extractable from the originals is also given
   under the name ``Input States.''}
\label{IHFIG}
\end{figure}

First, only copiers that do not use an ancillary subsystem $x$, entangled
with the copies, have been considered.
It is probably possible to obtain somewhat better performance
in ultimate information copying by using such helper subsystems, since discarding
$x$ after copying is completed partially relaxes the conditions that the copy states
$o$ and $c$ must satisfy to preserve unitarity (since one then  has more parameters left to
optimize over). It is not clear how much better one could do with such
helper states,
but we suspect not much better, since from Fig.~\ref{IHFIG}, the
copier considered here is only marginally better than several others
obtained by optimising over different indicators such as fidelity and
one-state copied information.

Secondly, for similar reasons, we have assumed that since both possible
input states $\op{\rho}^A_i$ are 
of equal purity \mbox{$\tr{(\op{\rho}^A_i)^2}$} (totally pure, in fact),
then both reduced copy states $\op{\rho}^B_i$ will be of equal purity also:
\begin{equation}
\tr{(\op{\rho}^B_1)^2} = \tr{(\op{\rho}^B_2)^2}.
\end{equation}
  This is also a property shared by all other copiers mentioned in this article.
The usual assumptions of Sec.~\ref{GPROP}, such as both copies being
equal, apply also.

So, an ancillaless copier, that produces two identical (usually imperfect) copies of any of two
possible pure signal states, that makes copies of the same purity whichever of
the two input states is sent, and that (given the above) maximizes the amount of
information that can be transmitted to each of the copies by any block-coding scheme
when the two input states are equiprobable,
 is given by the somewhat lengthy characterisation below.
 The details of how this was obtained have been left for Appendix~\ref{APP}.

There is a whole family of copying transformations, related by local unitary transformations
on the copies after they have stopped interacting with each other, which
give the same ultimate information copied $I_H^u$. Of these, we will
specify that particular one in this family which gives the greatest local fidelity between
the copies and originals.
The transformation can be written in terms of the parameters $r_m$ and $\phi_m$,
which have to be determined numerically. In terms of the initial states\ (\ref{instates}),
\begin{mathletters}
\begin{eqnarray}
\ket{\psi^A_1} &\to& \sqrt{\displaystyle\frac{1+r_m}{2}}\;\ket{b_1} +
\sqrt{\frac{1-r_m}{2}}\;\ket{b_2},
\end{eqnarray}

\begin{eqnarray}
\ket{\psi^A_2} &\to& \sqrt{\frac{x}{2}}\;\ket{b_1} + \sqrt{\frac{x}{2} - r_m\cos\phi_m}\;\,\ket{b_2}\nonumber\\
 &&\quad+ \sqrt{\frac{1-x+r_m\cos\phi_m}{2}}\;\Bigl(\;\ket{b_3}+\ket{b_4}\,\Bigr),
\end{eqnarray}
  where
\begin{equation}
  x = \half\left(1 + \cos^2\phi_m + 2 r_m\cos\phi_m + \sqrt{1-r_m^2}\,\sin^2\phi_m\right),
\end{equation}\end{mathletters}
  and the four  $\ket{b_j}$ are orthogonal basis states, given in 
 terms of the usual $\ket{+}$ and $\ket{-}$ basis states
  used in Eqs.\ (\ref{instates}) and\ (\ref{WZtransf}) by the matrix equation
\begin{equation}\label{UHO1}
\left(\begin{array}{c}\ket{b_1}\\\ket{b_2}\\\ket{b_3}\\\ket{b_4}\end{array}\right)
 = U
\left(\begin{array}{c}\ket{++}\\\ket{-+}\\\ket{+-}\\\ket{--}\end{array}\right),
\end{equation}
where the unitary matrix $U$ is
\end{multicols}

\begin{multicols}{1}
\widetext
\begin{equation}\label{UHO2}
U = \frac{1}{2}
\left(\begin{array}{cccc}1+\sin\phi_m/2&1-\sin\phi_m/2&\cos\phi_m/2&\cos\phi_m/2\\
1-\sin\phi_m/2&1+\sin\phi_m/2&-\cos\phi_m/2&-\cos\phi_m/2\\
-\cos\phi_m/2&\cos\phi_m/2&1+\sin\phi_m/2&\sin\phi_m/2-1\\
-\cos\phi_m/2&\cos\phi_m/2&\sin\phi_m/2-1&1+\sin\phi_m/2
\end{array}\right).
\end{equation}
\end{multicols}

\begin{multicols}{2}
As can be seen from the above, the basis states $\ket{b_j}$ are
entangled over the two copies.

The parameter $\phi_m$ is actually the angle between the Bloch vectors of
the two possible reduced copy states $\op{\rho}^B_i$, which can be written
\begin{mathletters}\begin{eqnarray}
\op{\rho}^B_1 &=& \frac{1}{2}\left(\begin{array}{cc}1+q&q_H\\q_H&1-q\end{array}\right),\\
\op{\rho}^B_2 &=& \frac{1}{2}\left(\begin{array}{cc}1-q&q_H\\q_H&1+q\end{array}\right),
\end{eqnarray}\end{mathletters}
where the parameters $q$ and $q_H$ are
\begin{mathletters}\begin{eqnarray}
q &=& r_m\sin\frac{\phi_m}{2}, \label{qU}\\
q_H &=& r_m\cos\frac{\phi_m}{2},
\end{eqnarray}\end{mathletters}
and appear in the expressions for $I_1$ and $I_H$.

Now $\cos\phi_m$ is dependent on $r_m$, and is given in terms of it as the
second largest\cite{root} real root of the following quartic polynomial in $\cos\phi_m$:
\end{multicols}
\pagebreak

\widetext
\begin{eqnarray}\label{ihpoly}
 0 &=& \cos^4\phi_m\left[r_m^2(2-r_m^2-2\sqrt{1-r_m^2})\right]
 +\cos^3\phi_m\left[4r_m^2(1-\sqrt{1-r_m^2})\right]\nonumber\\
 &&+\cos^2\phi_m\left\{2[r_m^4+2r_m^2+4f(\sqrt{1-r_m^2}-1)]\right\}
 +\cos\phi_m\left[4r_m^2(1+\sqrt{1-r_m^2}-4f)\right]\\
 &&+\left[(4f-1)^2-(1-r_m^2)^2+2(r_m^2-4f)\sqrt{1-r_m^2}\right].\nonumber
\end{eqnarray}

\begin{multicols}{2}
The ultimate copied information is given by
\begin{eqnarray}\label{IHbloch}
I_H^u &=&
  \half\left[ (1+r_m)\log_2(1+r_m) +
 (1-r_m)\log_2(1-r_m)\right]\nonumber\\
 &&-\half\left[ (1+q_H)\log_2(1+q_H) + (1-q_H)\log_2(1-q_H)\right],
\end{eqnarray}
%\narrowtext
which can be made a function of $r_m$ only, using Eq.\ (\ref{ihpoly}).
To obtain the
optimum copier, we find numerically the value of $r_m$ that maximizes
$I_H^u$ on \mbox{$r_m \in \left[\sqrt{1-f},1\right]$}. 

The one-state copied information $I_1^u$ is given by the same
expression in the distinguishability parameter $q$
as for the WZ copier [Eq.\ (\ref{WZIm})], with $q$ now given by Eq.\ (\ref{qU}).

\begin{figure}
\center{\epsfig{figure=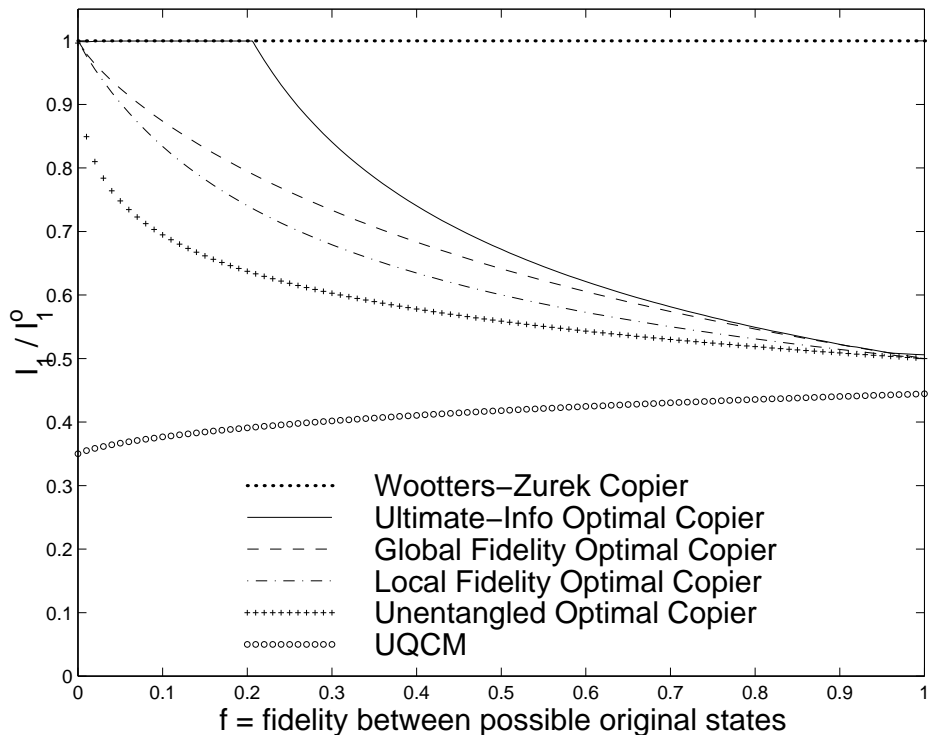,width=125mm}}
\caption{   One-state copied information (in bits per signal state) $I_1$ for the copying
   machines discussed in Secs.~\ref{3COP} and \ref{COMP} and Appendix~\ref{FAPP}, as a
   fraction of the maximum one-state information $I_1^o$ extractable from the
   input states\ (\ref{instates}), plotted as a function of the
   fidelity $f$ between the two pure input signal states.}
\label{IMFIG}
\end{figure}

It is interesting to note that, for input
states which are sufficiently
nonorthogonal (\mbox{$f \lesssim 0.206$}),
  the copier given here is just the WZ copier described in
Sec.~\ref{WZ}. In these cases, \mbox{$\phi_m = \pi$} and
\mbox{$r_m = \sqrt{1-f}$}. This sudden change in behavior (particularly
evident in Fig.~\ref{IMFIG} and
Fig.~\ref{FFIG} ) may be due to excluding the use of
ancillary subsystems. Allowing these may make the $I_H$ optimal
copier consistently better (although possibly not by much) than the
Wootters-Zurek for all values of $f$, even the small ones.

The local fidelity between copies and originals for this copier is
\begin{equation}
 F(\op{\rho}^A_i, \op{\rho}^B_i) =
  \half\left(1 + q\sqrt{1-f} + q_H\sqrt{f}\right).
\end{equation}

%========== NOENT =============================================
\subsection{An optimal copier that gives unentangled copies}
\label{NOENT}

As has been remarked by many previously, 
optimal quantum copiers typically produce highly entangled copies. This also
applies to the two quantum copiers given in 
Secs.~\ref{WZ} and~\ref{HOLEVO}. Nevertheless, copies
of some quality can be made without entanglement between them. This 
might be desirable in some situations.

Once again two simplifying assumptions have been made to make the
calculation easier.  It has been assumed that the copies are, again, unentangled with
ancillary machine states, and that the output state of the copier
is simply a product state of the two identical copies, rather than a
classical mixture of several such product states.  The 
case with additional machine states present might 
allow somewhat higher information transmission $I_H$ with block-coding
methods, for the same reasons as in Sec.~\ref{HOLEVO}. This would be interesting
to check, but we have not done this to date. Allowing classical correlations between
copies and a machine state subsystem $x$ does not, however,  improve information
transmission.

  Given the above two restrictions, a copier that optimizes
 both the  one-state and ultimate copied information, while keeping the copies unentangled, is given by
\begin{mathletters}\label{NEtransf}\begin{eqnarray}
\ket{\psi^A_1} &\to& 
  \frac{1+\sqrt{1-\sqrt{f}}}{2}\ket{++} + \frac{1-\sqrt{1-\sqrt{f}}}{2}\ket{--}\nonumber\\
 &&\qquad+ \half f^{1/4}\Bigl(\ket{+-}+\ket{-+}\Bigr), 
\end{eqnarray}
\begin{eqnarray}
\ket{\psi^A_2} &\to& 
\frac{1-\sqrt{1-\sqrt{f}}}{2}\ket{++} + \frac{ 1+\sqrt{1-\sqrt{f}}}{2}\ket{--}\nonumber\\
 &&\qquad+ \half f^{1/4}\Bigl(\ket{+-}+\ket{-+}\Bigr),
\end{eqnarray}\end{mathletters}
with notation identical to Eqs.\ (\ref{WZtransf}). See Appendix~\ref{NOEAPP}
for details of the optimisation.

  This gives pure state copies (they must be pure from the unitarity
  of the transformation, since the input states are pure, and the
  output state is \mbox{$\op{\rho}^B_i \otimes \op{\rho}^B_i$})
\begin{mathletters}\begin{eqnarray}
\op{\rho}^B_1 &=& \half\left(\begin{array}{cc}
  1+\sqrt{1-\sqrt{f}} & f^{1/4}\\
  f^{1/4} &  1-\sqrt{1-\sqrt{f}}\end{array}\right),\\
\op{\rho}^B_2 &=& \half\left(\begin{array}{cc}
  1-\sqrt{1-\sqrt{f}} & f^{1/4}\\
  f^{1/4} &  1+\sqrt{1-\sqrt{f}}\end{array}\right).
\end{eqnarray}\end{mathletters}
A family of copiers which do as well in the information
measures, but worse in local fidelity between originals and copies,
is given by making unitary transformations on the copies
individually. 

The one-state copied information $I_1^{\text{NE}}$ is given by the same expression in $q$
as for the WZ copier\ (\ref{WZIm}), with $q$ now given by
\begin{equation}
  q = \sqrt{1-\sqrt{f}}.
\end{equation}
The ultimate copied information  is
\begin{equation}
I_H^{\text{NE}} = 1 -
  \frac{1+f^{1/4}}{2}\log_2 (1+f^{1/4}) -
\frac{1-f^{1/4}}{2}\log_2 (1-f^{1/4}).
\end{equation}
The local fidelity of copies with respect to originals is
\begin{equation}
 F(\op{\rho}^A_i,\op{\rho}^B_i) = 
\half\left[ f^{3/4} + 1 +\sqrt{(1-f)(1-\sqrt{f})}\right].
\end{equation}
It turns out that this copier also gives the best local fidelity out
of such unentangling copiers (see Appendix~\ref{NOEAPP}).

%=============== COMPARISON ==============================================
\section{A Comparison of the Copiers}
\label{COMP}
  To see how well the copiers rate in terms of the information
  measures $I_H$ and $I_1$, we first need to determine how much information could be
  extracted from the input states if they were not copied. Since
the input states are not orthogonal for $f>0$, then a full bit of
  information cannot be extracted from each state even though they are
  equiprobable.

One finds that the information extractable one state at a time is 
\begin{equation}\label{Io}
I_1^o = \half\left[(1+q)\log_2(1+q) + (1-q)\log_2(1-q)\right],
\end{equation}
with the distinguishability parameter \mbox{$q = \sqrt{1-f}$}. This
is the same as with the Wootters-Zurek copier (\ref{WZIm}).
The ultimate information extractable from the signal if block-coding
methods are used is, however, unlike that for the WZ copier,  much larger:
\begin{equation}
I^o_H = 1 - \frac{1+\sqrt{f}}{2}\log_2(1+\!\sqrt{f}) -
\frac{1-\sqrt{f}}{2}\log_2(1-\!\sqrt{f}).
\end{equation}

It is interesting to compare the performance of the copiers given in
Sec.~\ref{3COP} to previously known fidelity-optimized ones. Three
will be considered here, and a brief summary of the copies they
produce is given in Appendix~\ref{FAPP} in terms of the input state
overlap parameter $f$.  

These three copiers are as follows.
(1) The universal quantum copying machine\cite{BuzekH:96} (UQCM), which
copies arbitrary qubits with a local fidelity of $5/6$. This is the maximum 
possible if it is to copy all with equal fidelity.
(2) A copier found by Bru\ss\ {\it et al.}\cite{Brussetal:98} that
optimizes 
the global fidelity when copying one of  two nonorthogonal input
states.
(3) A copier also found by Bru\ss\ {\it
et al.}\cite{Brussetal:98,Fuchs:98} that optimizes the local
fidelity when copying one of  two nonorthogonal input states.
So let us see how they compare in performance.

%======= IM ===================
\subsection{One-state copied information}
\label{COMPIM}
  The one-state copied information is a good indicator of the
efficiency of communicating classical data to the two copies. The
recovery and coding of the information in this case  relies only on measurement of
one-qubit states, and classical error-correction schemes. 

  Looking at Fig.~\ref{IMFIG}, one sees that the Wootters-Zurek
copier, apart from achieving the optimum and transmitting as much
one-state information to both copies as was encoded originally,  is
also far better at it than
any of the other copiers shown (except for the small-$f$ region,
where the ultimate-information optimized copier becomes the WZ).
The WZ copier has by far the simplest transformation out of these
copiers, so
 it seems that for basic information transmission the
simplest copier is the best.

  The fidelity-optimized copiers do not do  as
well as the WZ, which in itself is to be expected, as after all they were optimized
for fidelity, not information transfer. However, they do very much
worse, causing the loss of much information that could be regained if
better copiers were used. This shows quite clearly that fidelity is not
necessarily a good measure of the quality of the copies for all
situations. It is perhaps also surprising that even though we are
considering information transmitted to {\it one} copy here, 
the copier that has
been optimized for global fidelity between the combined output state and
perfect copies, does significantly better than the 
one that has been optimized
for local fidelity between a single copy and original. 

  The UQCM gives much less information transfer
than the other copiers, since all the others have been
specifically tailored for the two signal states, whereas the UQCM must
handle any arbitrary states with equal fidelity. 

  The copiers that give optimum unentangled copies do generally
significantly worse than the other copiers apart from the UQCM, but
one sees that all the copiers apart from the WZ copier and UQCM
converge to the same efficiency (much worse than the optimum) for high values of $f$, i.e., when the
signal states are not very orthogonal. 

  Note that a plot of the actual (rather than relative) amount of information extractable from the original
signal $I_1$ is shown in Fig.~\ref{IHFIG} as the Wootters-Zurek curve, since
$I_H^{\text{WZ}} = I_1^o$.

\subsection{Ultimate copied information}
\label{COMPIH}

The ultimate (Holevo bound) copied information $I_H$
gives an absolute maximum
on how much information could possibly be transmitted by a given
copier, with the best signaling scheme that is possible. In general, to
achieve this bound, the encoding/decoding scheme has to be very
elaborate, and it is often not achievable in practice due to
complexity. In the case of qubit systems being transmitted here, this
would entail making measurements of many-qubit observables to decode
the information: a difficult task at present. 

As can be seen in Fig.~\ref{IHFIG}, most of the
copiers cluster just below the optimal capacity achieved by the copier
of Sec.~\ref{HOLEVO}. While this is not necessarily the
absolute optimum that can be achieved, as there remains the
possibility that introducing helper machine states may increase this
bound, this bunching makes it seem plausible that no large gains can be
achieved beyond this. This ultimate-information  optimal
copier is quantitatively not much better than the Wootters-Zurek
copier. Its greatest gains, which are still quite modest, come  when
the overlap between signal states is high, where the absolute information
content in the signal is small. 

It can be seen that, while the no-cloning theorem did not stop one from
perfectly copying information contained in one state at a time, its
effect is strong where
block-coding schemes are allowed. This is because, if we restrict
ourselves to the one case at a time situation, we are not utilising
those properties of the states that are affected by the no-cloning theorem.
The difference between what can be extracted from a copy and from the
originals is quite striking, and for highly overlapping input states,
over
60\% of the information in the originals is unavailable from a copy.
 
The behavior of the copiers for high overlap between states is as one
would expect. That is,  the Wootters-Zurek copier becomes much less
efficient than the others when  block-coding schemes are used, as the other
copiers do not fully entangle the copies with each other, thus allowing
one to extract some extra information by looking at several sequential
states together. 

  Since the Wootters-Zurek copier has $I_1 = I_H$, by
comparing the values of $I_H$ for the local and global-fidelity-optimized copiers 
to the WZ copier, one can see that for these fidelity-optimal copiers,
much more information than $I_1$  can be sent to the copies by
allowing complicated block-coding schemes which use correlations
between subsequent signal states. This approach, however, is 
unhelpful with the Wootters-Zurek copier, and is of very little help
when using the 
the UQCM. 

  As for the other information measure, the global-fidelity-optimized copier does slightly better
than the local fidelity one. The unentangled copier does slightly
worse than the rest, except for the UQCM which is consistently  worse
on all counts, as it is not tailored to the input states like the others.

\subsection{Local fidelity}
\label{COMPF}
\begin{figure}
\center{\epsfig{figure=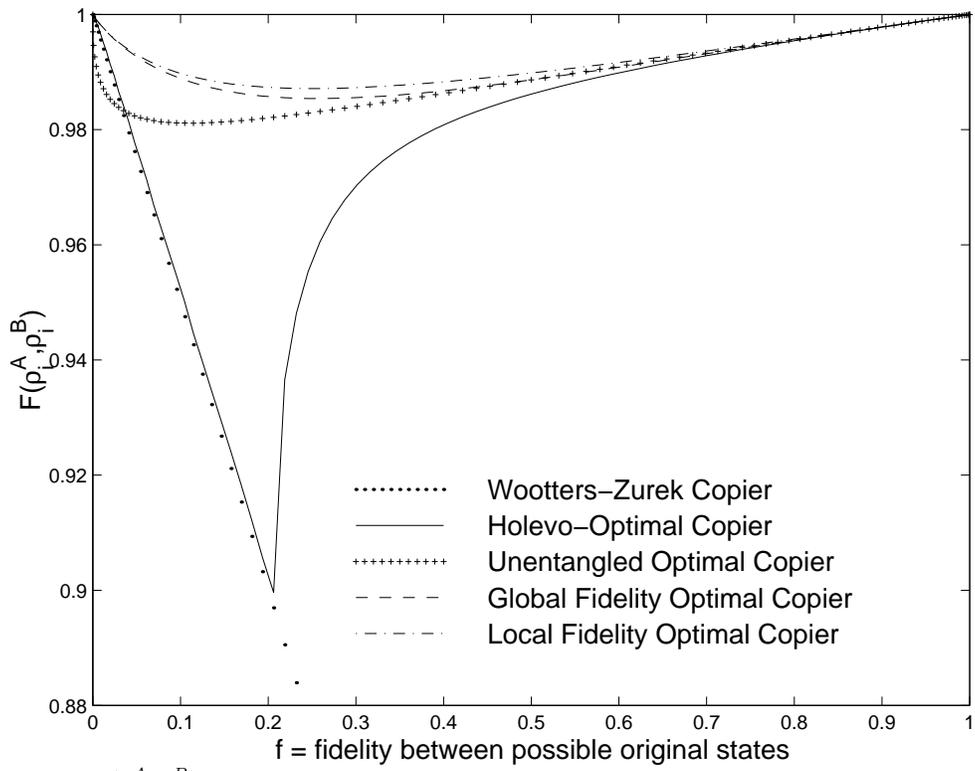,width=130mm}}
\caption{   Local fidelity \mbox{$F(\op{\rho}^A_i,\op{\rho}^B_i)$} between a copy
   and the original, for the copying
   machines discussed in Secs.~\ref{3COP} and \ref{COMP}
   and Appendix~\ref{FAPP}, as a function of $f$, the fidelity between the two
   input signal states.}
\label{FFIG}
\end{figure}

  This is shown for various copiers in Fig.~\ref{FFIG}.
The UQCM is absent
  from the plot, as its local fidelity lies far below the others shown
there.
Figures~\ref{IMFIG} and~\ref{FFIG} show quite clearly that fidelity
and information transfer quantify quite different properties of the
copying transformation, and one has to keep in mind which properties
are desired, before deciding on a quantity to characterize efficiency.

As expected, the best local fidelity occurs for the copier that was
  optimized for this, and the global fidelity optimal copier is almost
  as good. The WZ copier is no good at fidelity at all for
  significantly overlapping states. The unentangled copier is once
  again slightly worse than most of the others.
The sharp change in behavior for the ultimate-information optimal copier is 
particularly evident in this plot.

\section{Comments and Conclusions}
   It was seen in the previous section that quantum copiers optimized
for  fidelity measures are far from optimal for basic information
transmission to the copies, and, vice versa, information-optimized
copiers are far from optimized for fidelity between copies and
originals. This indicates that various measures of quality should be
used for quantum devices, depending on what final use is to be made of
the states created. 
  
Some other general trends that were seen for the quantum copying devices
that were considered, include the following.
   The ultimate-copied-information-optimized copier behaves more
similarly to the fidelity-optimized ones than to the one-state
optimized WZ copier (where it differs from the WZ). The
fidelity-optimized copiers are not bad when one allows multiparticle
measurements on the copies,
but are far from optimal if one does not.
This may be
because the fidelity-optimized copiers preserve some of the quantum
superposition of the input states (as evidenced by the off-diagonal
terms in the density matrices of the copies), whereas the WZ copier
makes the copies purely classical mixtures when they are considered
individually. To get extra information transmission by making
measurements on multistate observables, one needs some quantum
effects between the successive copy states, and these effects are
lacking with the WZ copier.

 A small, but perhaps surprising feature was that the
global-fidelity-optimized 
copier gave better performance  in the information measures
 than the local-fidelity-optimized one, even though only information
 flowing to one copy was considered. 
Other features seen include the poor performance of the UQCM relative
to the other copiers --- unsurprising, since the other ones are tailored
specifically to the two signal states, and the poorer performance when
the copies are made unentangled with each other.

For all copiers considered, when the input signal states are
 nonorthogonal,
 the information carrying capacity of a
channel between two observers is significantly greater when 
the receiver gets undisturbed states ($I_H^o$) than when the receiver
gets one copy, even when the copier is highly optimized
($I_H^u$). This is an information-theoretic manifestation of the
 no-cloning theorem.

%=============== APPENDIX ==============================================
\appendix
\section{Derivation of Ultimate-Information Optimal Copier}
\label{APP}
  The copier sought has the following properties: it takes one of two
\mbox{($i = 1,2$)} pure
input states\cite{dummy} $\op{\rho}^A_i$ of Hilbert space dimension 2,
and by a unitary transformation creates a state $\op{\rho}_i$ consisting of two
(possibly entangled) copies ($\op{\rho}^o_i$ and $\op{\rho}^c_i$),
again  of Hilbert space dimension 2. The state of
each copy, when the other copy is ignored, is identical, and both possible
copy states (corresponding to input states)
have equal purity, as measured by their self-fidelity $\tr{\op{\rho}^2}$.
  Assuming all states considered are normalized, these conditions can
be written as
\begin{eqnarray}\label{Hcnorm}
\text{normalization: }& \tr{\op{\rho}^A_i} = 1,\\\label{Hcinpure}
\text{input pure: }& \op{\rho}^A_i = \ket{\psi^A_i}\bra{\psi^A_i},\\\label{Hcuni}
\text{unitarity: }& \op{\rho}_i = \ket{\psi_i}\bra{\psi_i}, \\\label{Hcuni2}
& \tr{\op{\rho}_1\op{\rho}_2} = \tr{\op{\rho}^A_1\op{\rho}^A_2} = f,\\\label{Hcsym}
\text{symmetry: }& \op{\rho}^o_i = \text{Tr}_c[\op{\rho}_i] = \op{\rho}^c_i
    = \text{Tr}_o[\op{\rho}_i] =\op{\rho}^B_i,\\\label{Hcpure}
\text{equal purity: }& \tr{(\op{\rho}^B_1)^2} = \tr{(\op{\rho}^B_2)^2}.
\end{eqnarray}
 And, of course, on top of these conditions, the Holevo bound on
ultimate 
information copied $I_H$ is to be maximized.

The output states can be written in terms of a vector of complex expansion
coefficients in some basis as
\begin{equation}\label{HOout}
\ket{\psi_j} = \frac{1}{\sqrt{2}}\left[ \alpha_j, \beta_j
e^{i\phi_{\beta j}}, \gamma_j e^{i\phi_{\gamma j}}, \delta_j
e^{i\phi_{\delta j}} \right],
\end{equation}
where  \mbox{$\alpha_j$,$\beta_j$,$\gamma_j$,$\delta_j$ $\in [0, \sqrt{2}]$},
and the angles \mbox{$\phi_{\dots} \in [0, 2\pi)$}. Normalization gives
\mbox{$\alpha_j^2 + \beta_j^2 + \gamma_j^2 + \delta_j^2 = 2$}. One of the 
expansion coefficients can be made real and  positive, without
affecting the final bound, by multiplying by appropriate unphysical phase
factors, so let us do this to the $\alpha_j$.

Now, any two states in a two-dimensional Hilbert space (such as the
reduced states of the two possible copies $\op{\rho}^B_1$ and $\op{\rho}^B_2$),
can be described by two Bloch vectors ${\bf r}_i$. The states are then
given by
\begin{equation}\label{Blochdef}
  \op{\rho}_i({\bf r}_1) = \half\left(\op{I}+\bbox{\sigma}\cdot{\bf r}_i\right)
  \quad\text{where}\quad\bbox{\sigma} = \left[\op{\sigma}_1,\op{\sigma}_2,\op{\sigma}_3\right]
\end{equation}
and $\op{\sigma}_j$ are the Pauli matrices.
By an appropriate choice of basis, one of the two Bloch vectors can be
chosen to lie in an arbitrary direction, while the other is separated by
some angle $\phi_r$ from the first, both of them lying in a plane of our
choosing. Thus there are only three parameters for these two states that are not
arbitrary, depending on the choice of basis: the lengths of the Bloch vectors $r_i$,
and the angle between them $\phi_r$. Also, since
\begin{equation}
\tr{\op{\rho}_i({\bf r}_i)^2} = \half(1+|{\bf r}_i|^2),
\end{equation}
and we are assuming equal copy purity\ (\ref{Hcpure}), both Bloch vectors are of equal
length \mbox{$r=|{\bf r}_i|$}. Let us choose these Bloch vectors to be
\begin{equation}
{\bf r}_1 = r[0,0,1] \quad\text{and}\quad
{\bf r}_2 = r[\sin\phi_r,0,\cos\phi_r].
\end{equation}
Thus, without any loss of generality,
the copies can be written in an appropriate basis as
\begin{mathletters}\label{Hsta}\begin{eqnarray}
\op{\rho}^B_1 &=& \half\left(\begin{array}{cc}
1+r & 0\\ 0 & 1-r
\end{array}\right),\\
\op{\rho}^B_2 &=& \half\left(\begin{array}{cc}
1+r\cos\phi_r & r\sin\phi_r\\ r\sin\phi_r & 1-r\cos\phi_r
\end{array}\right).
\end{eqnarray}\end{mathletters}

Using Eqs.\ (\ref{Hsta}),\ (\ref{HOout}),  and conditions\
(\ref{Hcnorm}),\ (\ref{Hcsym}), one obtains the following restrictions on the
expansion coefficients of the total output states $\op{\rho}_i$:
\begin{mathletters}\label{HOcoefco}\begin{equation}\label{HOcoefcond}\begin{array}{rclcrcl}
\gamma_1 &=& \beta_1, &\ & \gamma_2 &=& \beta_2, \\
\beta_1^2 &=& 1+r-\alpha_1^2, &\ &
\beta_2^2 &=& 1+r\cos\phi_r-\alpha_2^2,\\
\delta_1^2 &=& \alpha_1^2-2r, &\ &
\delta_2^2 &=& \alpha_2^2-2r\cos\phi_r, 
\end{array}\end{equation}
\begin{eqnarray}
\label{HOb1coef}\beta_1\left(\alpha_1e^{i\phi_{\beta 1}} + \delta_1e^{\phi_{\delta
1}-\phi_{\beta 1}}\right) &=& 0,\\\label{HOb2coef}
\beta_2\left(\alpha_2e^{i\phi_{\beta 2}} + \delta_2e^{\phi_{\delta
2}-\phi_{\beta 2}}\right) &=& r\sin\phi_r.
\end{eqnarray}\end{mathletters}
Now Eq.\ (\ref{HOb1coef}) implies that either \mbox{$\beta_1 = 0$} or \mbox{($\alpha_1
= \delta_1$ and $2\phi_{\beta 1} = \phi_{\delta 1} + \pi$).} The second possibility
is uninteresting, as it
immediately leads to $r=0$, which gives $I_H = 0$ --- certainly not the
optimum case, one hopes!

Also, using the unitarity condition\ (\ref{Hcuni2}) and the equal purity
condition\ (\ref{Hcpure}), one obtains the restrictions
\begin{eqnarray}\label{HOf:}
2f &=& x+r(r-1)\cos\phi_r\nonumber\\
&&\quad+C\sqrt{1-r^2}\sqrt{x(x-2r\cos\phi_r)},
\end{eqnarray}\begin{eqnarray}
\label{HOn:}
r^2(1-\cos^2\phi_r) &=& 2(1+r\cos\phi_r-x)\\
&& \times[x-r\cos\phi_r+K\sqrt{x(x-2r\cos\phi_r)}],\nonumber
\end{eqnarray}
respectively. For brevity, the mutually independent parameters $x,K,C$
have been introduced, where
\begin{mathletters}\begin{eqnarray}
\label{Hpardef}
 x &=& \alpha_2^2, \\
  K &=& \cos(\phi_{\beta 2}+\phi_{\gamma 2} -
\phi_{\delta 2}), \\ C &=& \cos(\phi_{\gamma 2} - \phi_{\gamma 1}).
\end{eqnarray}\end{mathletters}
Note that the condition\ (\ref{HOn:}) is equivalent to Eq.\
(\ref{HOb2coef}).

Using Eqs.\ (\ref{Holevob}),\ (\ref{Blochdef}), and\ (\ref{Hsta}) leads to
$I_H$ being given by the expression
\begin{eqnarray}\label{IHexpr}
I_H &=& \half\left[ (1+r)\log_2(1+r) + (1-r)\log_2(1-r)\right] \\
     &&- \half\left[(1+q_H)\log_2(1+q_H) +
     (1-q_H)\log_2(1-q_H)\right],\nonumber
\end{eqnarray}
with
\begin{equation}
q_H = r\cos\frac{\phi_r}{2}.
\end{equation}

One finds that \mbox{$I_H(r,\cos\phi_r)$} is a monotonically decreasing
function of $\cos\phi_r$ --- thus, to maximize $I_H$ for a
given value of \mbox{$r=r_o$}, it suffices to minimize $\cos\phi_r$
  (i.e. make the angle between the possible copy Bloch vectors as close to
$\pi$ as possible). \mbox{$I_H(r,\cos\phi_r)$} is also a monotonically increasing
function of $r$.

For any particular values of $r$ and $\cos\phi_r$,
there are three parameters left to vary to try to satisfy Eqs.\ (\ref{HOf:})
and\ (\ref{HOn:}), after the relations\ (\ref{HOcoefco})
have been used: \mbox{$x$, $K$, and $C$}. Each of the two Eqs.\ (\ref{HOf:}),\
(\ref{HOn:}) will give an allowable range for $x$ (exactly which point in these
ranges is satisfied by the copier then depends on $C$ and $K$). The
ends of these ranges are given by
\begin{mathletters}\label{99}\begin{equation}
\frac{\partial\cos\phi_r}{\partial C} = 0 \quad\text{ or }\quad C = \pm 1
\end{equation}
for Eq.\ (\ref{HOf:}), and
\begin{equation}
\frac{\partial\cos\phi_r}{\partial K} = 0 \quad\text{ or }\quad K = \pm 1
\end{equation} 
\end{mathletters} for Eq.\ (\ref{HOn:}). Only those values of $\cos\phi_r$
for which the two $x$ ranges partially overlap give allowable copiers.
Now, for any particular \mbox{$r=r_o$}, if we vary
$\cos\phi_r$, the $x$ ranges will vary also. In particular, at that value of $\cos\phi_r$
which lies at the boundary of allowed \mbox{$\cos\phi_r(r_o)$} values, at least
one extremity of the first $x$ range, due to Eq.\ (\ref{HOf:}), will coincide with an extremity of the
second $x$ range due to Eq.\ (\ref{HOn:}). Of course, not all cases where $x$ range
extremities coincide will correspond to a $\cos\phi_r(r_o)$ extremity, but any parameters
for which such $x$ extremities coincide will give viable copiers [they could
be well within a region of allowed $\cos\phi_r(r_o)$ values]. Hence, if we look at all
the parameters [given by Eqs.\ (\ref{HOf:}),\ (\ref{HOn:}), and\
(\ref{99}) ] where
$x$ range extremities occur, then one of them will give the desired minimum $\cos\phi_r(r_o)$ value.
It turns out that this $\cos\phi_r(r_o)$ minimum corresponds to
\mbox{$K=C=1$} when \mbox{$r \in [\sqrt{1-f},1]$}. For \mbox{$r<\sqrt{1-f}$},
$\cos\phi_r(r_o)$ can reach its absolute minimum value of $-1$, but since $I_H$ is
also  monotonically increasing in $r$, the optimum $I_H$ copier must have
\mbox{$r \geqslant \sqrt{1-f}$}, so these low values of $r$ can be ignored.
This leads to the second largest real root of polynomial\ (\ref{ihpoly})
as the expression for $\cos\phi_r(r)$ that maximizes $I_H$ for a given \mbox{$r\geqslant\sqrt{1-f}$}.
The final value of $r$ that maximizes $I_H$ out of all the copiers
considered, $r_H$,
is given now by a straightforward, one-parameter maximization of
\mbox{$I_H{\bf (}r,\cos\phi_r(r){\bf )}$}
over \mbox{$r \in [\sqrt{1-f},1]$}. Because this calculation is simple, straightforward, and accurate numerically, but
not so simple analytically, an analytical solution has not been attempted.

Now, to find the particular transformation which, given input states\
(\ref{instates}), not only maximizes $I_H$ but also makes
the local copy-original fidelity as large as possible, first make the
Bloch vectors of the copies be in the same plane as the Bloch vectors of
the input states, and then make both pairs symmetric about a common axis.
The Bloch vectors of the input states are
\begin{equation}
{\bf s}_1 = [\sqrt{f},0,\sqrt{1-f}] \quad\text{and}\quad
{\bf s}_2 = [\sqrt{f},0,-\sqrt{1-f}].
\end{equation}
These are in the \mbox{$(\op{\sigma}_1 - \op{\sigma}_3)$} plane, and symmetrically
spaced about $[1,0,0]$. So, to achieve the desired optimum local fidelity copier,
the appropriate transformation of the input states
 is found to be
\begin{mathletters}
\begin{equation}
\ket{\psi^A_i} \to (U_H \otimes U_H)\ket{\psi_i},
\end{equation}
where $\ket{\psi_i}$ is given by Eq.\ (\ref{HOout}), and the unitary transformations
are
\begin{equation}
U_H = \left(\begin{array}{cc}
\cos\xi_H & \sin\xi_H\\ -\sin\xi_H & \cos\xi_H
\end{array}\right)
\quad\text{where}\quad
\xi_H = \frac{\phi_r(r_H) - \pi}{4}.
\end{equation}\end{mathletters}
  This can be written as Eqs.\ (\ref{UHO1}) and\ (\ref{UHO2}).

%========= NOENT deriv
\section{Derivation of Unentangled Optimal Copier}
\label{NOEAPP}
  Consider copiers producing product states of the copies. This transformation
  can be written
\begin{equation}\label{purenoent}
\op{\rho}^A_i \to \op{\rho}^B_i \otimes \op{\rho}^B_i \otimes \op{\rho}^x_i,
\end{equation}
where $\op{\rho}^B_i$ are the copies and $\op{\rho}^x_i$ is a helper machine
state. The only other constraint on the copier is that it must be unitary, which
means that traces are preserved. This immediately leads to $\op{\rho}^B_i$ and
$\op{\rho}^x_i$ being pure because the input states are pure (via $\tr{\op{\rho}^2}$).
Furthermore,
\begin{equation}
f = \left(\tr{\op{\rho}^B_1\op{\rho}^B_2}\right)^2 \tr{\op{\rho}^x_1\op{\rho}^x_2}
 = f_{12}^2 f_x,
\end{equation}
where $f_{12}$ and $f_x$ are the fidelities between, respectively,  the two copy and two machine states produced
  after input of originals. Thus, since \mbox{$f_x\leqslant1$}, it follows that
 \mbox{$\sqrt{f}\leqslant f_{12} \leqslant 1$}.

Let us start with optmizing for one-state information transfer $I_1$. It
is easily shown that for equiprobable input states, $I_1$ satisfies Eq.\
(\ref{WZIm}) with the distinguishability parameter given by
\begin{equation}
q = \sqrt{1-f_{12}}.
\end{equation}
  This is most straightforward to show using the Bloch vectors of the copies.
Since $I_1$ is  monotonically increasing with $q$, it will be maximized
when $q$ is maximized. This is when \mbox{$f_{12} = \sqrt{f}$}.

Now let us look at $I_H$. For qubit copy states, this is again given
by Eq.\ (\ref{IHexpr}),
and since the copies are pure, \mbox{$r=1$}, and one finds \mbox{$q_H = \sqrt{f_{12}}$}.
With \mbox{$r=1$}, $I_H$ depends only on $q_H$, and will reach
extreme values either when
\begin{equation}\label{ihqh}
\frac{d I_H}{d q_H} = \half\log_2 \left(\frac{1-q_H}{1+q_H}\right) = 0,
\end{equation}
or at the end points of the $q_H$ range: \mbox{$q_H = (f^{1/4} \text{ or } 1)$}.
One sees that Eq.\ (\ref{ihqh}) is only satisfied for \mbox{$q_H = f_{12} = f = 0$},
so for general $f$, extreme values of $I_H$ are reached at
\mbox{$f_{12} = 1$} or 
\mbox{$ f_{12}=\sqrt{f}$}.
\mbox{$f_{12} = 1$} leads to \mbox{$I_H=0$}, so the optimal value for $f_{12}$ is
again $\sqrt{f}$. Thus the same copiers that are optimal in $I_1$ are also optimal
in $I_H$.

Lastly, let us look at local fidelity. The fidelity between any two pure
states is given by
\begin{equation}
F(\op{\rho}_1,\op{\rho}_2) = \half\left(1+\cos\phi\right),
\end{equation}
in terms of $\phi$, the angle between their Bloch vectors.
To minimize the average over both possible inputs of this Bloch angle between originals and copies,
we choose the Bloch
vectors of the copies to lie in the same plane as the Bloch vectors of the originals,
and to be symmetric about the same axis. Obviously, in this case, the local fidelity will
be maximized if the Bloch angle between the copies is as similar to the Bloch angle
between the originals as possible (since the Bloch angle between original and copy
is half the difference between these). Since \mbox{$f_{12} = \sqrt{f} \geqslant f$}, this
means that we want $f_{12} = \sqrt{f}$ again.
Hence, the unentangled optimal copier given in Sec.~\ref{NOENT} is optimal
in all three indicators considered in this article.

Choosing Bloch vector parameters such that Eq.\ (\ref{purenoent}) holds,
\mbox{$f_{12} = \sqrt{f}$}, and local fidelity is optimized, easily leads
to the copier given in Eq.\ (\ref{NEtransf}).
It is simplest to use Bloch vectors for this calculation.

%===== F COPIERS =============================================
\section{Some Fidelity-Optimized Copiers}
\label{FAPP}

This section gives a brief summary of the fidelity-optimized copiers 
that are compared to the information-optimized ones in
Sec.~\ref{COMP}. Expressions are given in terms of $f$, the square
overlap between the two input states. Much more detail is given in the
literature.

%===glof
\subsection{The copier that optimizes the global fidelity}
\label{GLOF}

The quantum copying machine that optimizes the global fidelity
between the combined state of both copies and a state consisting of
unentangled perfect copies
has been found by Bru\ss\ {\it et al.}\cite{Brussetal:98} 
The copies produced are (with the help of a little algebra)
\begin{mathletters}\begin{eqnarray}
\op{\rho}^B_1 &=& \frac{1}{2}\left(\begin{array}{cc}
\ds 1 + \sqrt{\frac{1-f}{1+f}} &\ds \frac{f+\sqrt{f}}{1+f}\\
\ds\frac{f+\sqrt{f}}{1+f} &\ds 1 - \sqrt{\frac{1-f}{1+f}}
\end{array}\right),\\
\op{\rho}^B_2 &=& \frac{1}{2}\left(\begin{array}{cc}
\ds 1 - \sqrt{\frac{1-f}{1+f}} &\ds \frac{f+\sqrt{f}}{1+f}\\
\ds\frac{f+\sqrt{f}}{1+f} &\ds 1 + \sqrt{\frac{1-f}{1+f}}
\end{array}\right).
\end{eqnarray}\end{mathletters}
The local fidelity is [from Eq.\ (47) in Ref. \cite{Brussetal:98}]
\begin{equation}
F(\op{\rho}^A_i, \op{\rho}^B,i) = 
\half\left(1+\frac{(1-f)\sqrt{1+f}+f(1+\sqrt{f})}{1+f}\right),
\end{equation}
and the one-state copied information is given by Eq.\ (\ref{WZIm}) with 
distinguishability parameter
\begin{equation}
 q = \sqrt{\frac{1-f}{1+f}}.
\end{equation}
The ultimate copied information
is given by the expression\ (\ref{IHexpr}),
where $r$, the magnitude of the Bloch vectors of the copies, is in this case 
\begin{mathletters}
\begin{equation}
r = \frac{\sqrt{1+f(1+2\sqrt{f})}}{1+f},
\end{equation}
and the parameter $q_H$ is
\begin{equation}
q_H = \frac{f+\sqrt{f}}{1+f}.
\end{equation}\end{mathletters}

%=== lof
\subsection{The copier that optimizes the local fidelity}
\label{LOF}

 As in Appendix~\ref{GLOF}, Bru\ss\ {\it et al.} have found the
 copier that optimizes the local fidelity between a copy and
 the originals\cite{Brussetal:98,Fuchs:98}. 
  From Eqs. (C1)-(C6), and(C12) and subsequent discussion in Ref. \cite{Brussetal:98}, the copies are
 in the states
\begin{mathletters}\begin{eqnarray}
\op{\rho}^B_1 &=& \frac{\sec 2\phi}{2}\left(\begin{array}{cc}
\cos 2\phi+\sqrt{1-f} & (1+\sqrt{f})\sin 2\phi\\
(1+\sqrt{f})\sin 2\phi&\cos 2\phi-\sqrt{1-f} 
\end{array}\right),\\
\op{\rho}^B_2 &=& \frac{\sec 2\phi}{2}\left(\begin{array}{cc}
\cos 2\phi-\sqrt{1-f} & (1+\sqrt{f})\sin 2\phi\\
(1+\sqrt{f})\sin 2\phi&\cos 2\phi+\sqrt{1-f}, 
\end{array}\right),\end{eqnarray}
where the angle $\phi$ is defined by
\begin{equation}
\sin 2\phi = \frac{\sqrt{f}-1+\sqrt{1-2\sqrt{f}+9f}}{4\sqrt{f}}.
\end{equation}\end{mathletters}
The local fidelity is [rearranging Eq. (C11) of Ref.\cite{Brussetal:98}]
\begin{equation}
F(\op{\rho}^A_i,\op{\rho}^B_i) = 
\half\left\{ 1+\cos 2\phi[1-\!f + \sqrt{f}(1+\sqrt{f})\sin 2\phi]\right\}.
\end{equation}
After some algebra, one finds that 
\begin{mathletters}
\begin{eqnarray}
q &=& \sqrt{1-f}\cos 2\phi,\\
r &=& \cos 2\phi\sqrt{1-f+(1+\sqrt{f})^2\sin^2 2\phi},\\
q_H &=& \sin 2\phi \cos 2\phi (1+\sqrt{f}),
\end{eqnarray}\end{mathletters}
which can be used in expressions\ (\ref{WZIm}) and\ (\ref{IHexpr}),
respectively, to find
$I_1$ and $I_H$.

%===== uqcm
\subsection{The UQCM}
\label{UQCM}
The universal quantum copying machine \cite{BuzekH:96} 
copies any two-dimensional input states with an equal, optimal,
local fidelity of $5/6$. This copier is unique among those mentioned in this
article, in that it uses a machine helper state which becomes
entangled with both copies after the process is complete.
Given the input states\ (\ref{instates}) used in this article, the
UQCM will create the copies
\begin{mathletters}
\begin{eqnarray}
\op{\rho}^B_1 &=& \frac{1}{6}\left(\begin{array}{cc}
3 + \sqrt{1-f} & 2\sqrt{f} \\ 2\sqrt{f} & 3 - 2\sqrt{1-f} 
\end{array}\right),\\
\op{\rho}^B_1 &=& \frac{1}{6}\left(\begin{array}{cc}
3 - \sqrt{1-f} & 2\sqrt{f} \\ 2\sqrt{f} & 3 + 2\sqrt{1-f} 
\end{array}\right).
\end{eqnarray}\end{mathletters}

To calculate $I_1$ and $I_H$, use
\begin{mathletters}
\begin{eqnarray}
q &=& \frac{2}{3}\sqrt{1-f},\\
r &=& \frac{2}{3},\\
q_H &=& \frac{2}{3}\sqrt{f}
\end{eqnarray}
\end{mathletters}
in expressions\ (\ref{WZIm}) and\ (\ref{IHexpr}).

%======REFS=============================================

\end{multicols}

\begin{references}
\bibitem{POVMexplanation}
  A measurement scheme can in general be represented by a set of
positive operators $\{\op{A}_i\}$ which satisfy $\sum_i \op{A}_i =
\op{I}$, and the probability of obtaining measurement result $x_i$
from a state $\op{\rho}$ is
$\tr{\op{A}_i \op{\rho}}$ \cite{CavesD:94}.
\bibitem{root}
If $r_m=1$, use the largest real root, and if $r_m=0$, $\phi_m$ is of
no physical significance and thus can take any value, all of which lead to
an information content of zero, anyway.
\bibitem{dummy}
As well as  a ``dummy'' input 
state of Hilbert space dimension 2,  which is always in the same state
initially. This is needed to preserve unitarity of the transformation.
\end{references}
\end{document}